\documentclass[prd,preprint,nofootinbib]{revtex4-1}
\usepackage[latin1]{inputenc}
\usepackage[T1]{fontenc}
\usepackage{amsmath}
\usepackage{float}
\usepackage{graphicx}
\usepackage{comment}
\usepackage{caption}
\usepackage{subcaption}
\usepackage[usenames]{color}

\newcommand{\mat}[1]{\mbox{\boldmath{$#1$}}} 

\begin{document}


\title{Heavy quarkonium production in a strong magnetic field} 

\author{C.S.\ Machado}
\author{F.S.\ Navarra}
\author{E.G.\ de Oliveira}
\author{J. Noronha}
\affiliation{Instituto de Física, Universidade de São Paulo, C.P. 66318, 05315-970 São Paulo, SP, Brazil}
\author{M. Strickland}
\affiliation{Department of Physics, Kent State University, Kent, OH 44242, USA}

\date{\today}


\begin{abstract}
It is well known that in noncentral heavy-ion collisions a transient strong magnetic field is generated in the direction perpendicular to the  reaction 
plane. The maximal strength of this field is estimated to be $eB \sim m^2_{\pi} \sim 0.02~\text{GeV}^2$ at the RHIC and 
$eB \sim 15 m^2_{\pi} \sim 0.3~\text{GeV}^2$ at the LHC. We investigate the effects of a strong magnetic field on $B$ and $D$ mesons, focusing 
on the changes of the energy levels and the masses of the bound states. Using the Color Evaporation Model we discuss the possible changes 
in the production of  $J/\psi$ and $\Upsilon$.

\end{abstract}

\maketitle

\section{Introduction}

As  has been pointed out in  \cite{m1,m2}, a very strong magnetic field  is produced in  noncentral heavy-ion collisions in the direction 
perpendicular to the reaction plane. The maximum strength of the magnetic field is estimated to be $eB \sim m^2_{\pi} \sim 0.02~\text{GeV}^2$ at
 the RHIC ($\sqrt{s}=200~\text{GeV}$) and $eB \sim 15 m^2_{\pi} \sim 0.3~\text{GeV}^2$ at the LHC ($\sqrt{s}=4.5~\text{TeV}$). \footnote{We use 
natural units ($\hbar=c=1$) and the  conversion for the magnetic field is $\frac{1~\text{GeV}^2}{(\hbar c)^{3/2}}=1.44 
\times 10^{19}~\text{Gauss}$.}  This has several interesting phenomenological implications, which were discussed  
in detail in the recent review \cite{tuchin-13} (see also  \cite{t1,t2}). In this work we are interested in the effects of the magnetic fields on charm and 
bottom production.  Some of these effects have been already discussed in the literature.  In \cite{t1} a careful discussion of the evolution of a 
$J/\psi$ under the influence of a strong magnetic field was presented. In \cite{mu} the  magnetic conversion of $\eta_c$ into $J/\psi$ was considered. 
Here we investigate the effects of the  magnetic field on $B$ and $D$ mesons, focusing on the changes of the energy levels and of the masses of these 
bound states.  

In nucleus-nucleus collisions charm is produced mostly by gluon-gluon fusion. The produced $c - \bar{c}$ pair can have total spin equal to 0 or 1 and 
 all  up-down spin combinations are allowed. The pair is produced at a typical time of $t_c \simeq 1/  2 m_c \simeq  0.1$ fm. The magnetic field  is 
very strong in the  beginning of the collision, typically until  $t_B \simeq  0.2$ fm. Therefore it is reasonable to assume that charm production is 
strongly influenced by the magnetic field. The same argument applies to bottom production.  As soon as the charm quarks are produced they start to 
interact with each other and with the other quarks 
in the environment. After a while they form bound states, $D$'s or 
$J/\psi$'s. The nature of the quark-antiquark interaction depends on the nature of the surrounding matter. In very central collisions the medium is 
very likely a deconfined system of quarks and gluons, i.e.,  a hot quark-gluon plasma.  In the QGP the quark-antiquark potential is the one gluon 
exchange potential,  $V \propto  - 1/r$, which may additionally be affected by  color screening.  In these collisions the magnetic field is zero on average. 
When the impact parameter increases, the formed medium is less dense and the magnetic field becomes stronger.  In the limit of grazing collisions 
the magnetic field is very strong, the colliding system is  made essentially  of few nucleons  and there is no plasma. In this case the 
heavy quark- heavy antiquark potential may be well approximated by the Cornell potential $ V \propto - \alpha_s/r + \sigma r$, where $\alpha_s$ and 
$\sigma$ are constants. In \cite{haglin} it was shown that this potential reproduces also the gross features of heavy - light systems, such as the $D$ and 
$B$ mesons.   From the existing data on $x_F$ and rapidity distributions of  charm mesons, we know that they are produced with 
very low longitudinal momentum (low $x_F$  and small rapidity).  Therefore, from the point of view of an observer at the c.m.s.,  during the charm pair 
production, the magnetic field is  approximately constant in space and time and the pair moves in the field with low velocity and is not subject to a 
strong dissociating Lorentz force.  Moreover, since the internal velocity is small the system can be treated non-relativistically.  While different aspects 
of  charm production have already been addressed in other works, here we draw attention to the  interaction between the magnetic field and the spin of the
 quarks and the resulting changes in the  masses of the bound states.  

As it was pointed out in \cite{mu}, the spin-field coupling induces  $M_1$ transitions, 
converting spin zero into spin one states and vice-versa.  In a quantum field theory approach this corresponds to the absorption of a photon by a spin 
zero particle. In \cite{mu} the transition rate $\gamma + \eta_c \rightarrow J/\psi$ was estimated and found to be small, basically because it turned out 
to be proportional to $1/m_c^2$. In a classical approach this  process would correspond to  the energy transfer from the external magnetic field to 
$\eta_c$ meson.
Here we consider a similar type of transition, namely the process $\gamma + D \rightarrow D^*$.  In this case the transition amplitude is   proportional to 
$1/m_q^2$, where $m_q$ is the light quark mass.  In comparison with the previous case this probability is now much larger.  In the classical language, this
 corresponds to an energy transfer from the  $B$ field, which is the order of the pion mass and can thus induce the spin flip. 
In the heavy ion reaction considered here the strength of the magnetic field is of the order of $m_{\pi}$.  Since the M1 transition is important, in the 
presence of the strong magnetic field the pseudoscalar $D^0$ ($= 1/\sqrt{2} [ \mid \uparrow \downarrow \rangle - \mid\downarrow \uparrow \rangle] $) and the 
vector $D^{*0}$ ($= 1/\sqrt{2} [\mid\uparrow \downarrow \rangle + \mid\downarrow \uparrow \rangle] $)  become mixed.  At the same time the spin states 
($\mid\uparrow \downarrow \rangle $   and $ \mid\downarrow \uparrow \rangle $) have different masses, the former being much lighter than the latter. 
As it will be seen,  the magnetic field acts as a medium  in which the masses are different from the vacuum masses and some spin combinations have larger 
masses than others.  In this medium it becomes energetically favorable to produce  $\mid\uparrow \downarrow \rangle $, which has a mass that decreases 
with increasing magnetic 
field. It will later decay into $D^0$ and $D^{*0}$.  These considerations taken together suggest the following picture for  $D^0$ and $D^{*0}$ 
production, which we will explain in detail in this work:
First we produce  slowly moving $c$ and $ \bar{u}$ quarks with a certain spin combination, which do not yet form a meson and which interact immediately with 
each other through the potential $V$  and with the B field. Since the production occurs within the magnetic field some spin combinations 
($\mid\uparrow \downarrow \rangle$ and
$\mid\downarrow \downarrow \rangle$) are favored because they have smaller mass. These combinations are thus more easily produced and more abundant. 
In a dilute 
hadronic environment they interact through a Cornell type potential, evolve in the magnetic field, receiving energy from it, and eventually materialize as 
physical $D^0$ and $D^{*0}$ mesons. In a quark gluon plasma they interact with each other through the Coulomb potential and also  with the magnetic field and 
with other particles in the hot and dense medium. In many simulations they dissociate and then later recombine with other quarks during the hadronization
\cite{rec}.   

This text is organized as follows. In the next section we start with a simplified discussion based on semiclasssical arguments to determine qualitatively the 
dependence of the  charm  
meson masses on the magnetic field. In the subsequent section, we develop the non-relativistic quantum mechanics of the problem, following the textbook 
treatment 
given for the hydrogen atom in a magnetic field in \cite{Landau:1991} and  later refined in \cite{Karnakov:2003} and \cite{Machet,Vysotsky}. We adapt the 
formalism 
to a heavy and light quark system. In our approach the magnetic field is treated as an external constant field. We solve numerically the appropriate 
Schr\"odinger 
equation and compute the masses of the heavy bound states as a function of the magnetic field.   In section IV we use the obtained masses in the color 
evaporation model, 
to study $J/\psi$ and $\Upsilon$  production. As it  will be seen, these changes in the masses produce visible changes in the production cross sections. 
In section V we present a short summary and concluding remarks.

\section{A semiclassical approach}

Before discussing  numerical results, we would like to to gain more insight into the 
problem using a simplified semiclassical and analytical treatment, which  will be developed in what follows. 
The  Hamiltonian of a free particle under the action of the field can be obtained through the minimal substitution  
$\textbf{p} \rightarrow(\textbf{p} -e\textbf{A})$, where $ \textbf{A}$ is the vector potential, which, in cartesian coordinates and in the symmetric 
gauge is given by $\textbf{A}=\left(-By/2,Bx/2,0 \right)$ so that the magnetic field is oriented along the z-axis. With this choice we have 
$A^2 = B^2\rho^2/4$, where $\rho^2 = x^2 + y^2$. The full Hamiltonian is then obtained by including the other interactions:
\begin{equation}
\label{hamiltonian1}
\mathcal{H}=\frac{1}{2m}\left(\textbf{p} -e\textbf{A} \right)^2 - \mat{\mu} \cdot \textbf{B} + V(r), 
\end{equation}
where the second term represents  the spin coupling to the magnetic field ($\mat{\mu}$ is the intrinsic magnetic moment) and the third term
contains the  central potential, $V(r)$ ($r = \sqrt{\rho^2 + z^2}$). For quarks we have:
\begin{equation}
-\mat{\mu} \cdot \textbf{B} = - g\left(\frac{q}{2m}\right) \mat{s} \cdot   \textbf{B}  =- \frac{q B \sigma_z}{2m}  
\label{momag}
\end{equation}
where $q$ and $m$ are the quark charge and constituent mass respectively and  $\sigma_z= \pm 1$ is the spin projection along the $z$ direction.  
For simplicity we choose $\textbf{p} = p \hat{z}$ and hence $\textbf{p} \cdot \textbf{A} = 0$, which eliminates one cross term  in (\ref{hamiltonian1}). 
For a system of two particles interacting with each other and independently with the magnetic field (\ref{hamiltonian1}) can be immediately generalized to
\begin{equation}
\label{hamiltoniana2}
\mathcal{H}=\frac{p_1^2}{2m_1} + \frac{p_2^2}{2m_2} + \frac{(q_1 B)^2 \rho_1^2}{8m_1} + \frac{(q_2 B)^2 \rho_2^2}{8m_2}  
+ \mathcal{H}_s +  \lambda \frac{ \vec{\sigma_1} . \vec{\sigma_2}} {m_1 m_2} +   V(r),
\end{equation}
where
\begin{equation}
\mathcal{H}_s =  -  \frac{q_1 B \sigma_z^{(1)}}{2m_1} -  \frac{q_2 B\sigma_z^{(2)}}{2m_2}.
\label{hs}
\end{equation}
In the above expressions $r=|\vec{r_1} - \vec{r_2}|$ and $\vec{r_1}$ and $\vec{r_2}$ are the coordinates of the particles with respect 
to the center of mass, $\, \rho_1=\sqrt{x_1^2+y_1^2}$, $\rho_2=\sqrt{x_2^2+y_2^2}$  and $V(r)$ is the quark-antiquark 
central potential.  
In a strong magnetic field, the spin-spin interaction can be ignored.  We are considering the mesons $D$, $D^*$, $B$, and $B^*$, where (1) is the 
heavy quark and (2) is the light quark.  For hydrogen-like systems, we can use the approximation $m_1 \gg m_2$, $m_1 \gg p_1$,  $r_2 \gg r_1 \simeq 0$ 
and $\rho_2 \gg \rho_1 \simeq 0$.  Then, $r_2 = r$, $\rho_2 = \rho$, $p_2 = p$ and the reduced mass is  $\mu \simeq m_2$.  

In order to obtain a qualitative understanding and estimate the order of magnitude of the effect of the strong magnetic field on heavy quark 
bound states, we shall use the semiclassical approximation. We use the uncertainty relation $p\cdot r \sim  1$  to replace 
$p$ by  $1/r$. With these approximations (\ref{hamiltoniana2}) becomes: 
\begin{align}
\label{energia}
E(\rho)&=  \frac{1}{2\mu(\rho^2+z^2)} + \frac{(q_2 B)^2\rho^2}{8\mu}  - \frac{ q_2 B\sigma_z^{(2)}}{2\mu}  + V(\sqrt{\rho^2+z^2}), 
\end{align}
Since the magnetic field does not affect directly the motion along the $z$ direction, we shall, for simplicity fix the $z$ coordinate and choose $z=0$. 
Following the discussion in the introduction, we shall use the Cornell potential \cite{Eichten:1979ms} and its particular case,  the pure QCD Coulomb 
potential,  for the quark-antiquark interaction:
\begin{align}
\label{cornell}
V(\rho)&=-\frac{\kappa}{\rho} +\sigma \rho + C,
\end{align}
where $\kappa$ and $\sigma$ are the effective coupling and the string tension, that can be extracted from lattice calculations
and from  phenomenological analyses of heavy meson spectroscopy. While $\kappa$ and $\sigma$ are flavor independent, the constant $C$ is 
adjusted to  reproduce the mass of the lowest state of each heavy meson family. 
For ($c - \bar{u}$) we have $m_1=m_c$ with $q_1=2e/3$ and $m_2=m_{\bar{u}}$ with $q_2=-2e/3$. 
For ($b - \bar{d}$) we have $m_1=m_b$ with $q_1=-e/3$ and $m_2=m_{\bar{d}}$ with  $q_2=e/3$. 
Minimizing $E$ with respect to $\rho$ we find the equilibrium radius, $\rho_0$, and  the energy of the lowest bound state  $E(\rho_0)$. The mass of the 
system is given by:
\begin{equation}
\label{massa}
M_0 = m_1+m_2 + E(\rho_0).
\end{equation}
In order to finish our semiclassical calculation of the ground state energy, we would like to estimate the expectation value of  the spin term in 
(\ref{energia}) 
in $D^0$ states, i.e.,  $\langle D^0 | \mathcal{H}_s | D^0 \rangle$. However, as mentioned in the introduction 
this estimate is not well defined for pure $D^0$ or $D^{*0}$ states, since the spin-magnetic field interaction term changes the sign of the spin wave 
functions, 
converting $D^0$ into  $D^{*0}$ and vice-versa. Indeed, using the spin wave functions:
\begin{align}
|D^0 \rangle = \frac{1}{\sqrt{2}}\left(\mid\uparrow \downarrow  \rangle - \mid\downarrow \uparrow  \rangle \right). 
\label{d0}
\end{align}
and 
\begin{align}
|D^{*0} \rangle =\left\{\begin{array}{l}
\mid\uparrow \uparrow \rangle ; m_s=1  \\  \label{ds}
\frac{1}{\sqrt{2}}\left(\mid\uparrow \downarrow   \rangle + \mid\downarrow  \uparrow  \rangle \right); 
m_s=0 \\  \mid\downarrow \downarrow \rangle ; m_s=-1 
\end{array}\right. 
\end{align}
and the spin Hamiltonian (\ref{hs}) it  is easy to show that:
\begin{align}
\mathcal{H}_s|D^0 \rangle 
= \left( -\frac{ q_1 B}{2 m_1} -  \frac{q_2 B}{2 m_2} \right)  |D^{*0} \rangle 
\end{align}
The states $|D^0\rangle$ and $|D^{*0}\rangle$  are not eigenstates of $\mathcal{H}_s$. A  basis of eigenstates of $\mathcal{H}_s$ is given by 
$\mid\uparrow \uparrow \rangle$, $\mid\downarrow \uparrow \rangle$, $\mid\uparrow \downarrow \rangle$ and $\mid\downarrow \downarrow \rangle$. With these states, we can 
compute  expectation values of $\mathcal{H}_s$ with the approximations described above:
\begin{align}\label{eud} 
\langle \uparrow \downarrow |\mathcal{H}_s\mid\uparrow \downarrow  \rangle &= \langle \downarrow \downarrow |\mathcal{H}_s\mid\downarrow \downarrow  \rangle =
+ \frac{q_2 B}{2 m_2}, \\  \label{edu} 
\langle \downarrow \uparrow  |\mathcal{H}_s\mid\downarrow  \uparrow  \rangle &= \langle \uparrow \uparrow  |\mathcal{H}_s\mid\uparrow  \uparrow  \rangle =
- \frac{q_2 B}{2 m_2}, 
\end{align}
When we produce the  state $(c -  \bar{u})$, $q_2 = -2e/3$ and  (\ref{eud}) will lower the energy (\ref{energia}), while  (\ref{edu}) will raise it. This shift 
in the mass becomes more pronounced at higher values of the magnetic field. The expressions (\ref{eud}) (or (\ref{edu})) are inserted into (\ref{hs}), which is then (together with  (\ref{cornell})) inserted into (\ref{energia}). The latter is finally inserted into (\ref{massa}) to give the masses of the states as a function of the magnetic field. 
These functions are plotted  in Figs. 1 and 2, with  (\ref{massa})  normalized by the corresponding  vacuum masses. Analogous considerations hold for the 
$(b -\bar{d})$ states, which are also shown in the figures. Fig. 1 (a) shows  the favored spin combinations, while Fig. 1 (b) shows the disfavored ones.
As can be seen, the spin interaction can modify the mass of these states.
These results are in qualitative agreement with the relativistic calculation presented  in  \cite{simonov}  for light mesons. 

In the figures, we can see a significant mass change in the region of  the LHC ($eB \sim 15 m^2_{\pi} \sim 0.3~ \text{GeV}^2$), 
which can result in a change of the quarkonium produced cross section.
For strong magnetic fields  the states with higher masses are effectively suppressed. In what follows, we will study the lower energy states 
$\mid\downarrow \downarrow  \rangle$ and $\mid\uparrow \downarrow  \rangle$ in the case of the $(c-\bar{u})$ system.
In Fig. 1  we show the results for  the  $(c - \bar{u})$  and $(b - \bar{d})$ systems  with the Cornell potential and in Fig. 2, with the QCD Coulomb potential.

\begin{figure}
\begin{center}
\begin{subfigure}{.5\textwidth}
    \includegraphics[width=.9\linewidth]{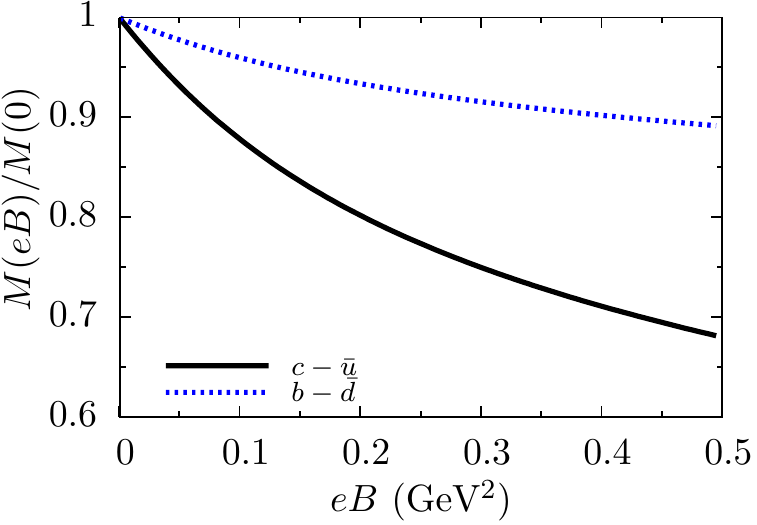}
    \caption{}
  \label{fig:sc}
\end{subfigure}%
\begin{subfigure}{.5\textwidth}
   \includegraphics[width=.9\linewidth]{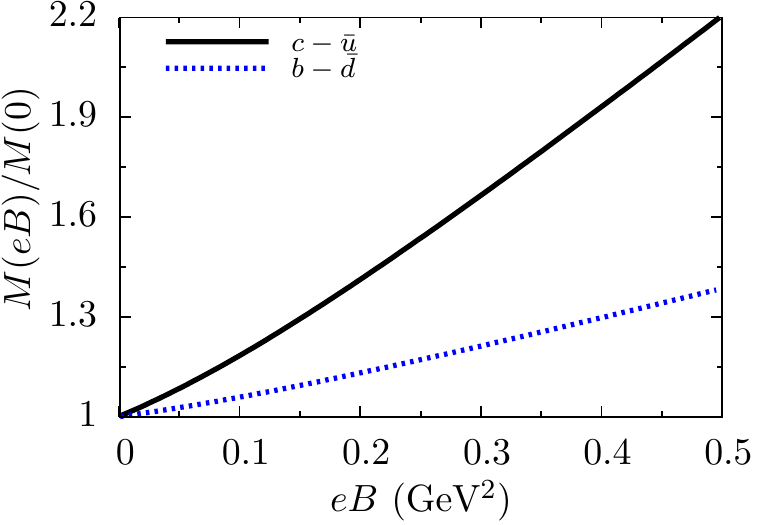}
   \caption{}
  \label{fig:sc2}
\end{subfigure}
\end{center}
\caption{Masses of the $q \bar{q}$ systems interacting through the Cornell potential as a function of the magnetic field.  
(a) $(c - \bar{u})$ ($\mid\uparrow \downarrow\rangle$) and $(b - \bar{d})$ ($\mid\downarrow \uparrow \rangle$).  
(b) $(c-\bar{u})$ ($\mid\downarrow \uparrow \rangle$) and $b-\bar{d}$ ($\mid\uparrow \downarrow\rangle$).}
\label{fig:test}
\end{figure}

\begin{figure}
\begin{center}
\begin{subfigure}{.5\textwidth}
  \includegraphics[width=.9\linewidth]{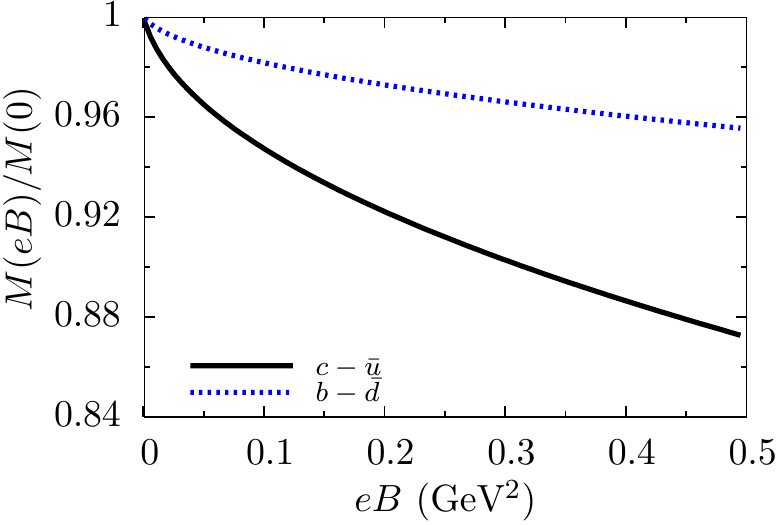}
  \caption{}
  \label{fig:sc3}
\end{subfigure}%
\begin{subfigure}{.5\textwidth}
  \includegraphics[width=.9\linewidth]{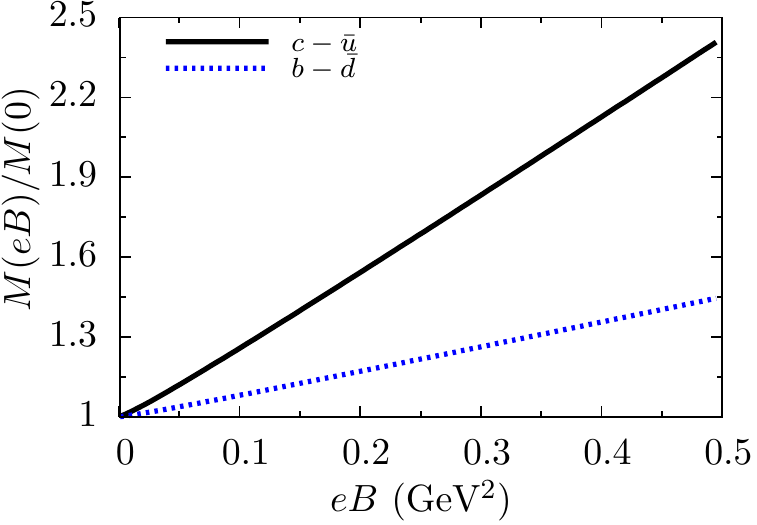}
  \caption{}
  \label{fig:sc4}
\end{subfigure}
\caption{Masses of $q \bar{q}$ systems interacting through the QCD Coulomb potential as a function of  $eB$.  
(a) $(c - \bar{u})$ ($\mid\uparrow \downarrow\rangle$) and $(b - \bar{d})$ ($\mid\downarrow \uparrow \rangle$).  
(b) $(c-\bar{u})$ ($\mid\downarrow \uparrow \rangle$) and $b-\bar{d}$ ($\mid\uparrow \downarrow\rangle$).}
\label{fig:test2}
\end{center}
\end{figure}

With this  simple model we predict  that the (rising or falling)  bevavior of the mass with the magnetic field can be attributed to the 
spin.  As will be seen in the next section, this  prediction is also in remarkable quantitative agreement with the numerical solution of 
the Schr\"odinger  equation.   This simple behavior of the bound state mass might change if the light quark mass would be significantly changed 
by the magnetic field. Indeed, in \cite{shov} it was shown that the magnetic field induces the generation of a dynamical mass for the light quarks, 
which turns out to be always smaller than the constituent quark mass used in potential model calculations, such as the one presented here. Therefore, 
we shall treat the constituent quark mass as a constant.

We could extend our model to the study of the $J/\psi$. However, in this case the "hydrogen-like" approximation is no longer valid, 
since both the quark and the antiquark have the same mass. The potential is no longer central (due to the lack of factorization of center of mass and 
internal motion) and the algebra becomes a bit more cumbersome.  A simple
estimate can be made by replacing the mass $m_2$ (or $\mu$) by the reduced mass of the $c - \bar{c}$ system, which is now much larger and hence
suppresses the spin effects. As expected, the change in the bound state mass is less than 5\% and will be neglected in what follows. 

\section{Numerical solution}

In this section we solve the Schr\"odinger equation for the Hamiltonian (\ref{hamiltoniana2}) with the same approximations used before, except for the momentum, 
which is now the standard momentum operator.  In cylindrical coordinates the vector potential has components 
$A_{\phi}= B\rho/2$, $A_{\rho}=A_z=0$  and the magnetic field is in the $z$ direction.  The Hamiltonian can then be written as:
\begin{equation} 
\mathcal{H} = -\frac{1}{2m}\nabla^2 + \frac{q²}{2m}\left(\frac{B\rho}{2} \right)^2 -\frac{qB}{4mi}\frac{\partial}{\partial\phi} - \frac{q B \sigma_z}{2m}.  
\end{equation}
The Schr\"odinger equation is given by:
\begin{equation}
\label{eq2d}
-\frac{1}{2m}\left[\frac{1}{\rho}\frac{\partial}{\partial \rho}\left(\rho \frac{\partial \psi}{\partial \rho} \right)+\frac{1}{\rho^ 2}
\frac{\partial^2 \psi}{\partial \phi^2}+\frac{\partial^2 \psi}{\partial z^2} \right]
- \frac{1}{2}i\omega_H\frac{\partial \psi}{\partial \phi}+\frac{1}{8}m\omega_H^2\rho^2\psi - \frac{qB}{2m}\sigma_z\psi = (E-V)\psi,
\end{equation}
where $m$ is the light quark  mass and $\omega_H = |qB|/m$. We can make the following  Ansatz for the wave function:
\begin{equation}
\label{wf}
\psi(\rho,z,\phi)=\chi(\rho,z)e^{im_{\phi}\phi}.
\end{equation} 
Considering only the ground state of the system,  which is azimuthally symmetric, we have $m_{\phi}=0$. 
We then insert (\ref{wf}) into (\ref{eq2d}) to find:
\begin{align}
-\frac{1}{2m}\left[\frac{1}{\rho}\frac{\partial}{\partial \rho}\left(\rho \frac{\partial \chi}{\partial \rho} \right)+
\frac{\partial^2 \chi}{\partial z^2} \right] +\frac{1}{8}m\omega_H^2\rho^2\chi - \frac{qB}{2m}\sigma_z\chi= (E-V(\rho,z))\chi
\end{align}
We solve the above equation numerically with a method described briefly in Appendix A. We assume a constant magnetic field and we use the Cornell 
potential with 
the parameters chosen so as to reproduce the experimental masses of the ($c - \bar{u}$) and ($ b - \bar{d}$) systems  in vacuum ($e B = 0$). They are: 
$m_c=1.37~GeV$, $m_b=4.79~\text{GeV}$, $m_{u,d}=0.20~\text{GeV}$, $\kappa=0.506$ and $\sigma=0.1695~\text{GeV}^2$. Moreover we need to use the constants
$C=-0.516~\text{GeV}$ for  ($b - \bar{d}$)  and $C=-0.544~\text{GeV}$ for   ($c - \bar{u}$) to obtain the measured values  $m_{B^{0}}=  5279.50\pm 0.30~\text{MeV}$ and  
$m_{D^{0}}= 1864.80\pm 0.14~\text{MeV}$. 
As mentioned in the introduction the Cornell potential should be relevant for less central collisions.  For the more central 
ones we expect that the produced pair will interact in a deconfined medium and hence the quark-antiquark potential can be approximated by a QCD screened 
Coulomb potential.  As a check of our numerical method we will compare results obtained with a pure Coulomb potential  
with the analytical predictions resulting from the Karnakov-Popov equation \cite{Karnakov:2003}. 
This equation gives the energy levels for the hydrogen atom in a strong magnetic field and it  can be  adapted  for mesons with a light and a heavy quark, as
discussed in  Appendix B. The numerical results for the mass (scaled by the vacuum value) as a function of the magnetic field  can be seen in Fig. 3. 

\begin{figure}
\begin{center}
\begin{subfigure}{.5\textwidth}
  \includegraphics[width=.9\linewidth]{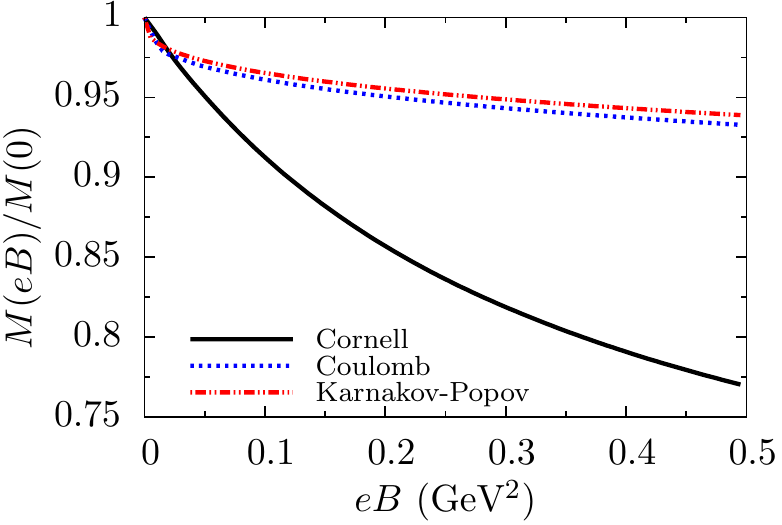}
  \caption{}
  \label{dratio}
\end{subfigure}%
\begin{subfigure}{.5\textwidth}
  \includegraphics[width=.9\linewidth]{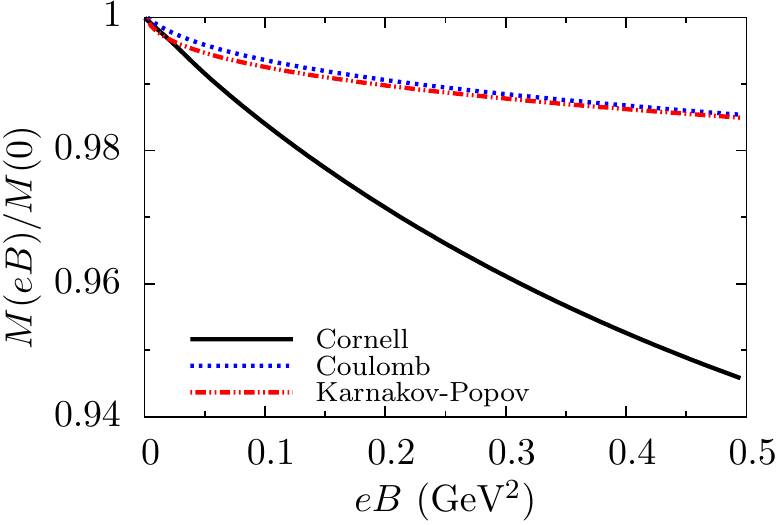}
  \caption{}
  \label{bratio}
\end{subfigure}
\begin{flushleft}
\caption{Masses of the quark-antiquark systems  as a function of the magnetic field. Numerical results obtained with the Cornell potential 
(solid lines), with the QCD Coulomb potential (dotted lines) and the analytical results obtained with the Karnakov-Popov equation (dotted-dashed lines).
(a) $(c - \bar{u})$ ($\mid\uparrow \downarrow\rangle$). (b) $(b - \bar{d})$ ($\mid\downarrow \uparrow \rangle$).}
\end{flushleft}
\label{fig:test3}
\end{center}
\end{figure}
The theoretical uncertainties of these results can be estimated considering the errors in the main parameters of the calculation, which are the reduced mass, 
$\mu$ (which is essentially the light quark constituent mass), the strong coupling, $\kappa$, and the string tension $\sigma$. Using the  current values  
found in the literature, we varied each parameter between a maximum and a minimum  keeping the others fixed. 
The results are shown in Fig. 4. From the figures we can conclude that the mass reduction effect is very robust and the amount of reduction may change 
by up to 15 \% for different parameter choices. 
\begin{figure}
\begin{center}
\begin{subfigure}{.5\textwidth}
  \includegraphics[width=.9\linewidth]{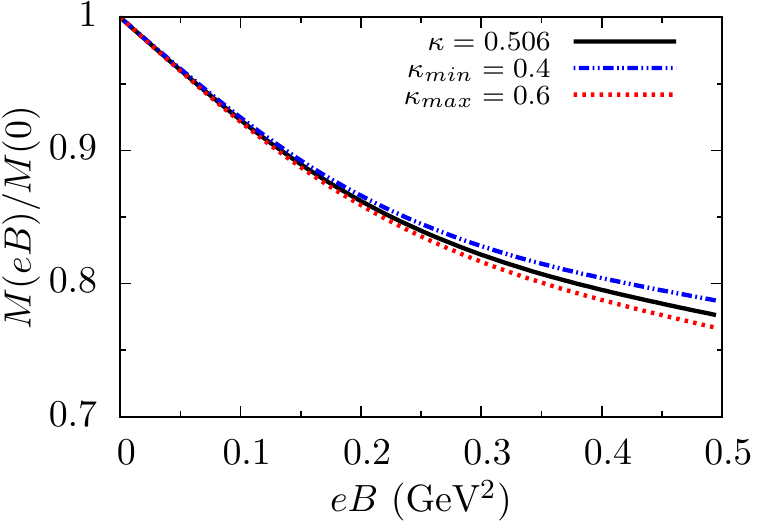}
  \caption{}
  \label{kappa}
\end{subfigure}%
\begin{subfigure}{.5\textwidth}
  \includegraphics[width=.9\linewidth]{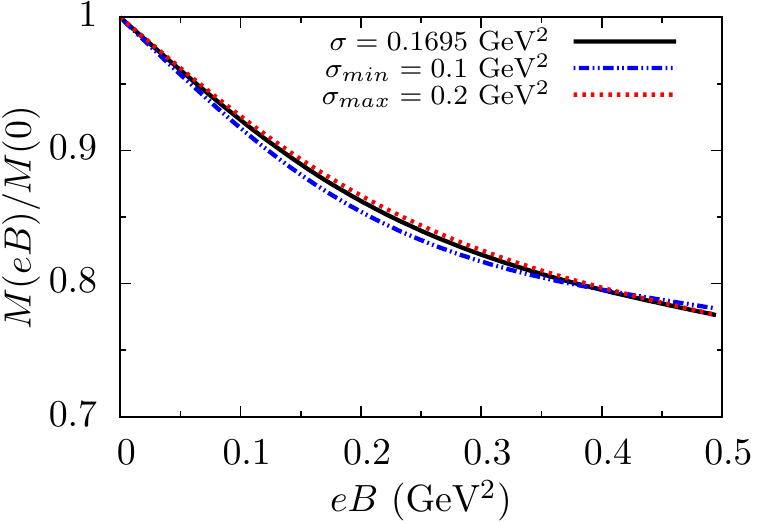}
  \caption{}
  \label{sigma}
\end{subfigure}
\label{fig:test3.5}
\begin{subfigure}{.5\textwidth}
  \includegraphics[width=.9\linewidth]{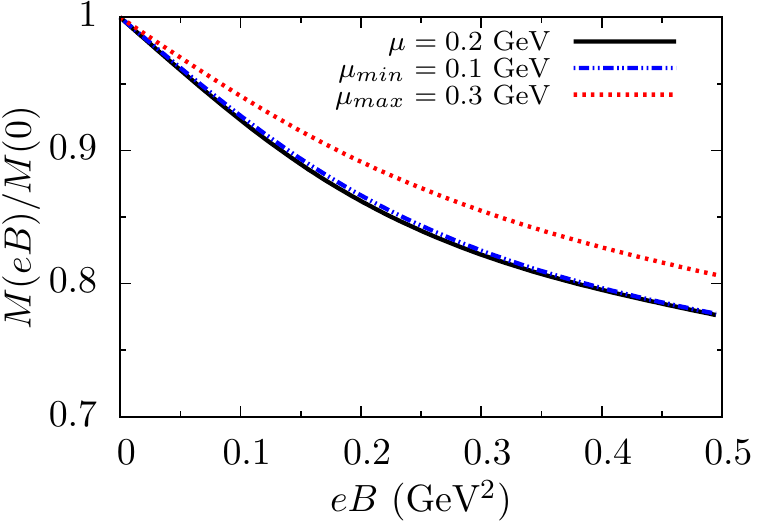}
  \caption{}
  \label{mu}
\end{subfigure}
\caption{Mass of the  ($c - \bar{u}$)  ($\mid\uparrow  \downarrow \rangle$) system as a function of  the magnetic field. 
Numerical results for the Cornell potential, with three different values of (a) the strong coupling, $\kappa$, 
(b) the string tension, $\sigma$ and (c) the reduced mass, $\mu$.}
\label{fig:test4}
\end{center}
\end{figure}
We close this section emphasizing that we observe a mass reduction in the ($c - \bar{u}$)  ($\mid \uparrow \downarrow \rangle$) states, both in analytical 
and numerical calculations. Once formed these states may evolve to form $D$ or $D^*$ mesons. This reduction is not observed in $c - \bar{c}$ states.  
The mass change in the open charm states may affect the production of hidden charm ($J/\psi$).  While, in the presence of a magnetic field, is remains
equally difficult to produce a $J/\psi$,  it becomes easier to produce a $D - \bar{D}$ pairs, which now have smaller masses. This idea can be implemented in 
a straightforward way with the help of the color evaporation model (CEM), in which the $D$ mass appears explicitly.  Similar considerations hold for the 
bottom sector. In the next section we investigate quantitatively how these mass changes modify the $J/\psi$ and $\Upsilon$ production cross sections. 

\section{ Heavy quarkonium  production in the color evaporation model}

The color evaporation model is very popular  \cite{cem1,cem2,cem3} and enjoys a great phenomenological success. 
Nowadays calculations with this model  can be found in textbooks \cite{book}. Nevertheless we shall,
in what follows, give some formulas to introduce the notation and to stress the role played by the $D$ mesons in the production of
$J/\psi$. For the sake of definiteness we will study the production of charmonium. The extension of the formulas to bottomonium is straightforward and  
numerical results for both cases will be presented.  

In the CEM, charmonium is defined kinematically as a $c - \bar{c}$ state with mass below the $D - \bar{D}$ threshold, i.e.,  
$(2 m_c)^2 < m^2 < (2 m_D)^2 $. 
At leading order (LO) the cross section is computed with the use of perturbative QCD for the diagrams of the elementary processes
$q + \bar{q} \rightarrow c + \bar{c}$ and $ g + g  \rightarrow c + \bar{c}$ convoluted with the parton densities in the projectile and in the target. 
The production cross section of a $c - \bar{c}$  pair with invariant mass $m$ is given by: 
\begin{align}
\label{producao2}
\frac{d\sigma_{c\bar{c}}}{dx_{F} dm^2} &= \int_0^1 dx_1 dx_2 \delta(x_1 x_2 s - m^2 )\delta(x_F - x_1 + x_2) H_{AB}\left(x_1, x_2; m^2\right) \nonumber\\
& = \frac{1}{s\sqrt{x^2_F + 4m^2/s}} H_{AB}\left(x_{01}, x_{02}; m^2\right),
\end{align}
with,
\begin{align}
x_{01;02} = \frac{1}{2}\left(\pm x_F + \sqrt{x_F^2 + 4m^2/s} \right),
\end{align}
where $x_F$ is the fractional momentum of the produced pair and $\sqrt{s}$ is the COM energy of a nucleon-nucleon collision. The $H_{AB}$ function is given by:
\begin{align}
\label{hfunction}
 H_{AB}\left(x_1, x_2; \mu^2\right)= \sum_{q=u,d,s}\left[f^A_q(x_1,\mu^2)f^B_{\bar{q}}(x_2,\mu^2)+f^A_{\bar{q}}(x_1,\mu^2)f^B_q(x_2,\mu^2) 
\right]\sigma_{q\bar{q}}(\mu^2) & \nonumber \\
+ f^A_g(x_1,\mu^2)f^B_g(x_2,\mu^2)\sigma_{gg}(\mu^2), &
\end{align}
and is computed at the scale $ \mu^2 = m^2 = x_1 x_2 s$. The functions $f_q$, $f_{\bar{q}}$ and $f_g$ are the quark, antiquark, and gluon distribution functions in
the proton, which we take from the CTEQ parametrizations \cite{cteq}. 
The leading order cross sections in terms of the pair invariant mass are given by:
\begin{align}
\sigma_{gg}(m^2) = \frac{\pi \alpha_s(m^2)}{3m^2}\left\{\left(1+\frac{4m_c^2}{m^2}+\frac{m_c^4}{m^4}\ln \left[\frac{1+\lambda}{1-\lambda} \right] 
\right) -\frac{1}{4}\left(7 + \frac{31m_c^2}{m^2} \right)\lambda \right\},
\end{align}
\begin{align}
\sigma_{q\bar{q}}(m^2) = \frac{8\pi\alpha_s^2(m^2)}{27m^2}\left( 1+\frac{2m_c^2}{m^2}\right)\lambda,
\end{align}
where $\lambda=\sqrt{1-4m_c^2/m^2}$ and $m_c$ is the $c$ quark  mass. The production cross section of the charmonium state is given by:
\begin{equation}
\frac{d\sigma_{J/\psi}}{d x_F} = F_{J/\psi}\int_{4m_c^2}^{4m_{D^0}^2} dm^2 \frac{d\sigma_{c\bar{c}}}{dx_F dm^2}.
\label{massint}
\end{equation}
In the previous section we showed that the strong magnetic field created in a heavy ion collision can modify the mass of a meson with a heavy and a 
light quark. The reduction of the $D$ ($B$) meson mass can consequently modify the $J/\psi$ ($\Upsilon$) production cross section since this mass squared  
enters  in the upper limit of the integral. According to the results shown in Fig. 3 with an uncertainty of $\simeq 5 \%$, we assume a reduction of 
$5\%$, $15\%$ and $25\%$ in the mass of the  $D$ and a reduction of  $3\%$, $5\%$ and $8\%$ in the mass  of the $B$ meson.  In Figs. 5 and 6 we  show the 
effect of this reduction in the $x_F$ distribution for $\sqrt{s}=4.5~\text{TeV}$  and in Fig. 7 we show the effect in the total cross section. 
The CEM parameters are  $F_{J/\psi}=0.025$ and  $F_{\Upsilon}=0.046$ \cite{book}.
\begin{figure}
\begin{center}
\begin{subfigure}{.5\textwidth}
  \includegraphics[width=.9\linewidth]{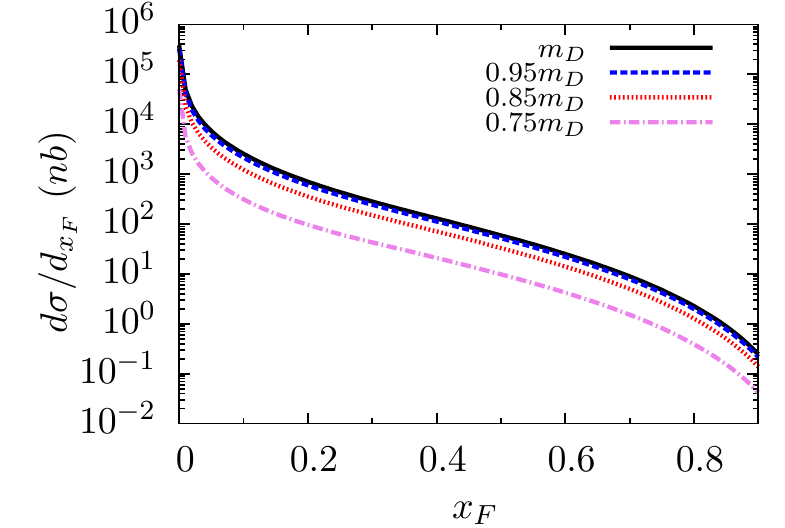}
  \caption{}
  \label{fig:cem}
\end{subfigure}%
\begin{subfigure}{.5\textwidth}
  \includegraphics[width=.9\linewidth]{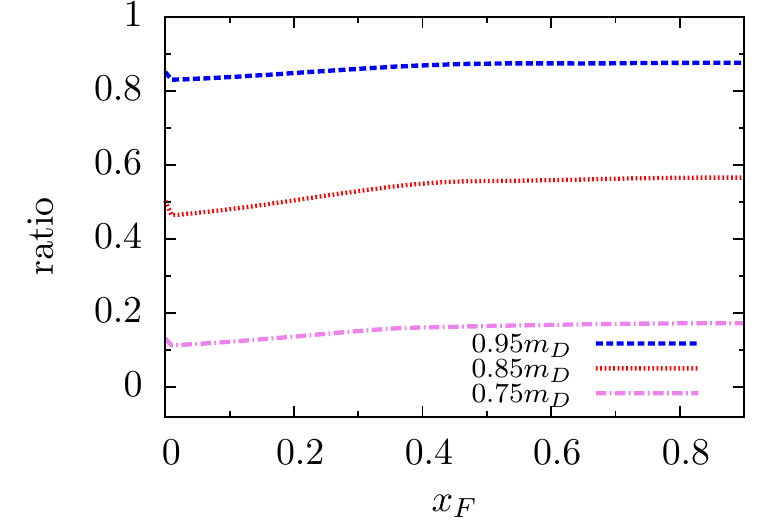}
  \caption{}
  \label{fig:ratioD}
\end{subfigure}
\caption{(a) Differential $J/\psi$ production cross section at $\sqrt{s} = 4.5~\text{TeV}$ for several values of the $D$ meson  mass. 
(b) Differential cross sections normalized by calculation performed with the vacuum ($eB=0$) $D$ mass.}
\label{fig:test5}
\end{center}
\end{figure}
\begin{figure}
\begin{center}
\begin{subfigure}{.5\textwidth} 
  \includegraphics[width=.9\linewidth]{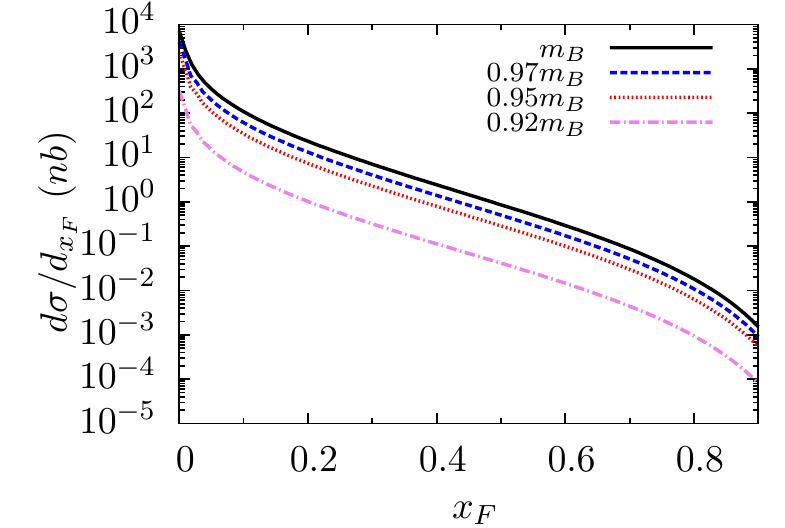}
  \caption{}
  \label{fig:cemB}
\end{subfigure}%
\begin{subfigure}{.5\textwidth}
  \includegraphics[width=.9\linewidth]{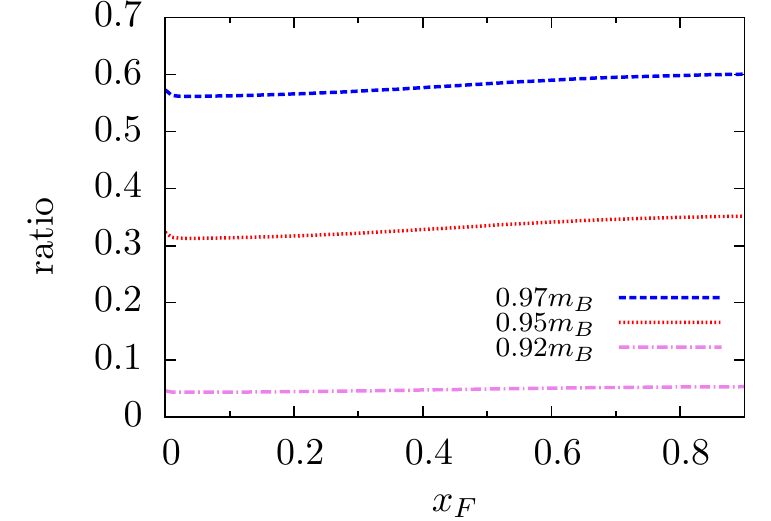}
  \caption{}
  \label{fig:ratioB}
\end{subfigure}
\caption{(a) Differential $\Upsilon$ production cross section at $\sqrt{s} = 4.5~\text{TeV}$ for several values of the $B$ meson  mass. 
(b) Differential cross sections normalized by calculation performed with the vacuum ($eB=0$) $B$ mass.}
\label{fig:test6}
\end{center}
\end{figure}
\begin{figure}
\begin{center}
\begin{subfigure}{.5\textwidth}
  \includegraphics[width=.9\linewidth]{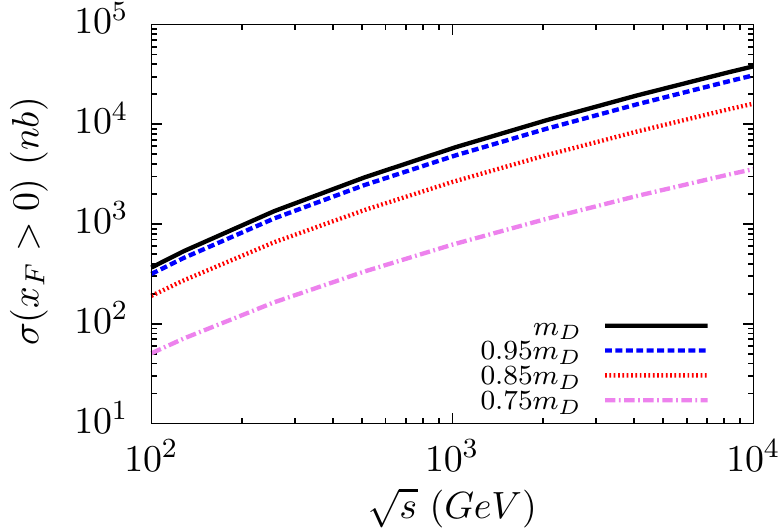}
  \caption{}
  \label{fig:cemtotalD}
\end{subfigure}%
\begin{subfigure}{.5\textwidth}
  \includegraphics[width=.9\linewidth]{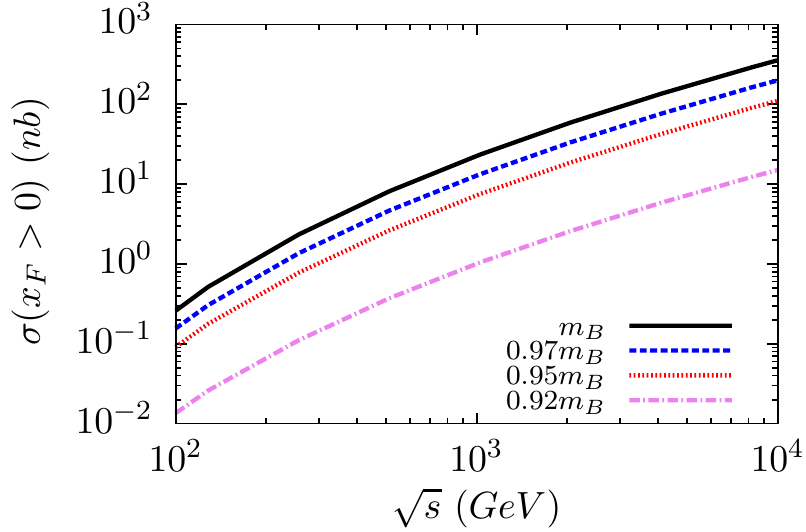}
  \caption{}
  \label{fig:cemtotalB}
\end{subfigure}
\caption{(a) Total cross section of (a) $J/\psi$ and (b) $\Upsilon$ production as a function of the c.m.s. energy for several values of the $D$ and $B$ masses.}
\label{fig:test7}
\end{center}
\end{figure}
We can observe that the mass change caused by the magnetic field can reduce the total cross section and the $x_F$ distribution by almost one order of magnitude.  
Moreover, the reduction in the $\Upsilon$ production cross section is, even for a modest reduction in $m_B$, of the same order of magnitude as the reduction in the  
$J/\psi$ production cross section. At first sight, in view of  (\ref{momag}) and (\ref{hs}), this might seem surprising.  However from (\ref{massint}) we see that  
the integration domain on $m^2$  grows with the mass difference between the heavy quark and the corresponding heavy meson, $m_M - m_Q$. In the case of the bottom a 
reduction of only 9\% in the $B$ mass ($0.09 \times 5.279 = 0.475$ GeV )  is almost enough to close the mass window $m_B - m_b = 0.49$ GeV and reduce the cross 
section to zero! Of course, at this stage our result should be taken just as an indication.  In order to check this effect we would need to simulate the whole 
collision process including all 
the other aspects that are known to affect particle production, such as nuclear shadowing, parton saturation, hydrodynamical flow (in central collisions), 
rescattering, 
final state interactions, etc. The point we wish to emphasize is that including the magnetic effect in these simulations seems to be mandatory.  
Having made this 
cautionary  remark, even though it might be premature, it is tempting to take a look on $J/\psi$ production data in heavy ion collisions.  Based on the 
calculations 
presented here we would expect that in the most non-central collisions there would be some extra suppression of $J/\psi$ due to open charm mass reduction. 
Indeed, in the 
centrality dependence in the data \cite{phenix} and also the analysis performed in \cite{levin}, we notice that for  $J/\psi$ with  
$y=0$, low $p_T$ ($1 < p_T < 3$ GeV)  
and in the least central collisions ($40 - 92$\%) there is a suppression stronger than expected. Since this is precisely the region where the effect
 discussed here 
should be most important, we feel encouraged to further develop and refine the present calculation.  

\section{Summary}

The main purpose of this work was to explore the effect of the magnetic field  on heavy meson production in heavy ion collisions. 
With the guidance of a  semi-classical model we could anticipate that the field induces a  mass reduction of  hydrogen-like heavy 
bound states, an effect  which is mainly due to the coupling between the spin and the magnetic field. Then, in a more realistic calculation, 
we numerically solved the three-dimensional Schr\"odinger equation with the  Cornell and the QCD Coulomb potential and confirmed the expectation of the analysis 
performed with the semi-classical model.  The obtained mass reduction turned out to be non-negligible. We explored the implication of the $D$ 
(and $B$) mass reduction for the $J/\psi$ (and $\Upsilon$) production using the Color Evaporation Model. The effect was surprisingly large both for  
$J/\psi$ and for $\Upsilon$. We made use of several simplifying assumptions herein.  Given these assumptions we found that 
the effect on $J/\psi$ production seems to be large enough to justify the inclusion of  magnetic field effects on detailed simulations of heavy ion collisions.


\begin{acknowledgments}
This work has been partly supported by CNPq and FAPESP. E.G.O.\ is supported by
FAPESP  under contract 2012/05469-4 and C.S.M.\ is supported by FAPESP  under contract 2011/05619-3. The authors are deeply grateful to R. D. Matheus, 
R. Higa, and L.G. Ferreira Filho.
\end{acknowledgments}

\appendix

\section{SPECTRAL METHOD}

The spectral method approximates the solution of partial differential equations as a linear combination of continuous and periodic functions. 
We will describe the method for a one-dimensional equation. The extension to the higher dimensional cases is straightforward. 
Let us consider the equation
\begin{equation}
- \frac{\mathrm{d}^2}{\mathrm{d}x ^2} f(x) + x^2 f(x) = E f(x)
\label{difa1}
\end{equation}
The  solutions can be written in terms of sines
\begin{equation}
f(x) = \sum_{j=1}^\infty A_j \sin\left[ \frac{j \pi}{2 L} (x+L) \right].
\label{serifa1}
\end{equation}
This solution is only valid in the range $-L < x < L$.
Inserting  (\ref{serifa1}) into (\ref{difa1}) we find
\begin{equation}
\label{spectral}
\sum_{j=1}^\infty A_j \left( \frac{j \pi}{2 L} \right)^2 \sin\left[ \frac{j \pi}{2 L} (x+L) \right] + 
x^2 \sum_{j=1}^\infty A_j \sin\left[ \frac{j \pi}{2 L} (x+L) \right] = E \sum_{j=1}^\infty A_j \sin\left[ \frac{j \pi}{2 L} (x+L) \right].
\end{equation}
Multiplying the above equation by the $k^{\rm th}$ term of the series and integrating over $x$ we have 
\begin{align}
\sum_{j=1}^\infty A_j \left( \frac{j \pi}{2 L} \right)^2 \int_{-L}^L \mathrm{d} x \sin\left[ \frac{k \pi}{2 L} (x+L) \right] 
\sin\left[ \frac{j \pi}{2 L} (x+L) \right] \\ 
+ \sum_{j=1}^\infty A_j  \int_{-L}^L \mathrm{d} x \sin\left[ \frac{k \pi}{2 L} (x+L) \right] x^2 \sin\left[ \frac{j \pi}{2 L} (x+L) \right] \\ 
= E \sum_{j=1}^\infty A_j \int_{-L}^L \mathrm{d} x \sin\left[ \frac{k \pi}{2 L} (x+L) \right] \sin\left[ \frac{j \pi}{2 L} (x+L) \right].
\end{align}
We can use the ortogonality of sine functions to  rewrite  (\ref{spectral}) as
\begin{equation}
\int_{-L}^L \mathrm{d} x \sin\left[ \frac{j \pi}{2 L} (x+L) \right] \sin\left[ \frac{k \pi}{2 L} (x+L) \right] = \delta_{jk} L.
\end{equation}
Therefore,
\begin{equation}
A_k \left( \frac{k \pi}{2 L} \right)^2 + \sum_{j=1}^\infty A_j C_{jk} = E A_k,
\end{equation}
where the coefficient $C_{jk}$ is
\begin{equation}
C_{jk} = \frac{1}{L}
\int_{-L}^L \mathrm{d} x \sin\left[ \frac{k \pi}{2 L} (x+L) \right] x^2 \sin\left[ \frac{j \pi}{2 L} (x+L) \right].
\end{equation}
In order to improve computational efficiency, we rewrite the equation using the sum of cosines
\begin{equation}
C_{jk} = \frac{1}{2L}
\int_{-L}^L \mathrm{d} x x^2 \left\{ \cos\left[ \frac{(k-j) \pi}{2 L} (x+L) \right] - \cos\left[ \frac{(j+k) \pi}{2 L} (x+L) \right] \right\}.
\end{equation}
Now we have to solve numerically the follow eigenvalue equation to obtain de energy
\begin{equation}
A_k \left( \frac{k \pi}{2 L} \right)^2 + \sum_{j=1}^\infty A_j \frac{1}{2L}
\int_{-L}^L \mathrm{d} x x^2 \left\{ \cos\left[ \frac{(k-j) \pi}{2 L} (x+L) \right] - \cos\left[ \frac{(j+k) \pi}{2 L} (x+L) \right] \right\} = E A_k.
\end{equation}


\section{Rescaling the  KP  equation}

The Karnakov-Popov equation gives the change of the energy levels of the hydrogen atom due to the magnetic field 
(details can be found in Ref.~\cite{Karnakov:2003}). For the ground state $m=0$ we have
\footnote{We use the expansion $\psi(1-1/\lambda)\approx\psi(1)-\pi^2/(3\lambda)$ valid for strong magnetic field i.e $\lambda \rightarrow \infty$.}
\begin{equation}
\ln\left(\frac{eB}{m_e^2 e^4 }\right) = \lambda + 2\ln\lambda - \psi(1) - \frac{\pi^2}{3\lambda} + \ln 2,
\end{equation}
where
\begin{align}
E=-\frac{m_e e^4}{2\hbar^2} \lambda^2,
\end{align}
and $\psi$ is the derivative of the gamma function. For the $B$ and $D$ mesons bound by a Coulomb potential, $V(r) = -\kappa/r $, we have  
\begin{equation}
\ln(qB) - \ln(\mu^2 \kappa^2) = \lambda + 2\ln\lambda - \psi(1) - \frac{\pi^2}{3\lambda}  + \ln 2,
\end{equation}
where  $\mu$ is the reduced mass of the system and
\begin{align}
E=-\frac{\mu \kappa^2}{2} \lambda^2.
\end{align}



\end{document}